\documentclass[prl,twocolumn,showpacs]{revtex4}
\usepackage{graphicx}
\begin{document}

\title{Frustrated spin model as a
hard-sphere liquid}
\author{M. V. Mostovoy, D. I. Khomskii, and J. Knoester}
\affiliation{Materials Science Center,
University of Groningen, Nijenborgh 4, 9747 AG Groningen,
The Netherlands}

\author{N. V. Prokof'ev}
\affiliation{Physics and Astronomy, Hasbrouck Laboratory,
University of Massachusetts, Amherst, MA 01003, USA}

\begin{abstract}
We show that one-dimensional topological objects (kinks) are
natural degrees of freedom for an antiferromagnetic Ising model on
a triangular lattice. Its ground states and the coexistence of
spin ordering with an extensive zero-temperature entropy can be
easily understood in terms of kinks forming a hard-sphere liquid.
Using this picture we explain effects of quantum spin dynamics on
that frustrated model, which we also study numerically.
\end{abstract}

\date{\today}


\maketitle

\looseness = -1 Geometrically frustrated materials
recently emerged as a new broad class of solids with
interesting and rather unusual properties
\cite{SchifferRamirez}. While some of these systems stay
disordered at all temperatures, others order often in an
unexpected way, showing no universality typical for
critical behavior of conventional systems. One may wonder,
however, to which extent the complexity of ground states
and excitations of frustrated models is their genuine
property and to which it is, essentially, a conventional
behavior obscured by an unfortunate choice of variables,
in which these models are formulated.  In this Letter we
consider a frustrated Ising model showing a
spin-density-wave (SDW) ordering, more common for systems
with continuous symmetries. We explain the origin of this
strange behavior and give a simple description of ground
states of that model (the number of which grows
exponentially with the system size) by mapping it on a
hard-sphere liquid. We use this approach to study the role
of quantum spin fluctuations, which for frustrated systems
is a challenging theoretical problem
\cite{Waldtmann,CanalsLacroix}. We also show that the
low-energy states of our model form a large number of
valleys separated by energy barriers, which prevents the
system from reaching thermal equilibrium at low
temperatures, but does not result in a spin-glass behavior
in the absence of quenched disorder
\cite{AeppliChandra,Gardner}.

\looseness = -1 One of the simplest classical frustrated models
describes the Ising spins $\sigma_i = \pm 1$ on sites of a
triangular lattice with nearest-neighbor antiferromagnetic
interactions:
\begin{equation}
E = J \sum_{\langle i,j\rangle} \sigma_i \sigma_j, \;\;\;
J > 0. \label{eq:classical}
\end{equation}
This system stays disordered at all nonzero temperatures
\cite{Wannier,Houtappel}, while at $T = 0$ the spins order
periodically
\begin{equation}
\langle \sigma_i \sigma_j \rangle \propto
\frac{1}{r_{ij}^\eta} \cos \frac{2 \pi r_{ij}}{3},
\label{eq:correlator}
\end{equation}
where $r_{ij}$ is the distance between the spins and $\eta
= 1/2$ \cite{Stephenson}. This algebraic order coexists
with an extensive zero temperature entropy $S_0 \approx
0.323 k_B$ per spin \cite{Wannier,Houtappel}.

\looseness = -1 This entropically induced ordering
resembles the crystallization in hard-sphere liquids at
high volume fractions \cite{Alder,Weber}. We show below
that the model Eq.(\ref{eq:classical}) can indeed be
mapped on a two-dimensional liquid of topological domain
walls (kinks), the motion of which is confined to one
spatial dimension. This mapping also provides useful
insights into the physics of the quantum version of
Eq.(\ref{eq:classical})
\begin{equation}
H = J \sum_{\langle i,j\rangle} \sigma^z_i \sigma^z_j - h
\sum_i \sigma_i^x, \label{eq:quantum}
\end{equation}
where $\sigma^{x,z}_i$ are the Pauli matrices and $h$ is
the transverse field (the classical  and quantum
frustrated models Eq.(\ref{eq:classical}) and
Eq.(\ref{eq:quantum}) are referred to below as,
respectively, CFM and QFM). The critical behavior in the
QFM was recently discussed by Moessner {\em et al.}, who
argued that the ordering of quantum spins is long-ranged
at low temperatures and that the ordered and disordered
phases are separated by a phase with algebraically
decaying spin correlations
\cite{MoessnerSondhiChandra,MoessnerSondhi}. Here we show
that the difference in the critical behavior of the CFM
and QFM originates from a higher rigidity of the quantum
kink crystal. We also perform numerical simulations of the
QFM and find an unexpected specific heat anomaly at strong
transverse fields.

\noindent{\em Kinks in frustrated Ising model:} We shall consider
the triangular lattice as an array of coupled chains, running in
the $x$-direction. As neighboring chains are shifted with respect
to each other, we will distinguish even and odd chains. In each
chain we perform the transformation from spins to kinks, which are
domain walls separating two different Neel states and carrying the
topological charge $q = \pm1$ (kinks and antikinks). The chain
energy equals $2 J N$, where $N$ is the number of kinks,
independent of kink positions. The interchain spin coupling gives
rise to interactions between pairs of kinks in neighboring chains
with the potential
\begin{equation}
V(x_o - x_e) = 2 J q_o q_e \mbox{sign}(x_o - x_e),
\label{eq:Vint}
\end{equation}
where $q_o$($q_e$) and $x_o$($x_e$) are, respectively, the
topological charges and $x$-coordinates of the kinks in
the odd(even) chain. This potential only depends on the
sign of the relative coordinate $x_o - x_e$. Therefore,
the energy of the CFM is completely determined by the
number of kinks and their relative ordering in neighboring
chains. These loose interactions between kinks are  much
different from those in unfrustrated models (e.g., the
Ising model on a square lattice) which grow linearly with
the distance between kinks,  confining them into pairs.

\begin{figure}[tbp]
\centering
\includegraphics[width=7cm]{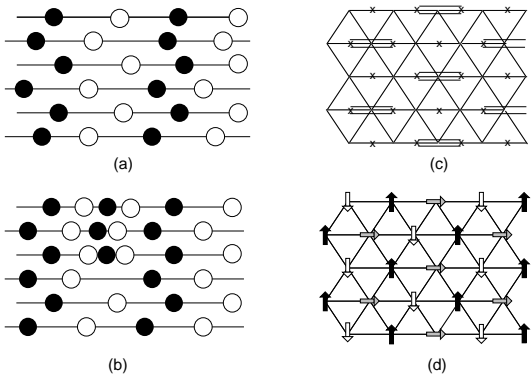}
\caption{{\protect\footnotesize (a) Optimal relative ordering of
kinks; the black(white) circles correspond to kinks with the
topological charge $+1(-1)$; (b) Dislocation in the kink crystal
with the energy $2J$;(c) The kink crystal, in which the kinks
delocalized over pairs of neighboring sites are indicated by
dimers; (d) The spin ordering corresponding to the kink crystal.}}
\label{fig:order}
\end{figure}

\looseness = -1 While in unfrustrated models the density of kinks
vanishes at low temperatures, in the CFM kinks are present in
ground states. In the energetically most favorable relative
ordering the $q = +1$ kink in an odd chain has the
nearest-neighbor $q = +1(-1)$ kinks in two neighboring even chains
from the right/left and vice versa (see Fig.~\ref{fig:order}a), in
which case the kink creation energy $2J$ is exactly compensated by
the energy of its interactions with kinks in neighboring chains,
i.e. in the ground states kinks cost no energy. For the favorable
relative ordering the number of kinks $N$ in each chain has to be
the same. Therefore, the ground states form distinct classes
labeled by $N$. Since shifts of kinks that preserve the ordering
do not change energy, each class still contains a large number of
states, resulting in an extensive ground state entropy.

\looseness = -1 To describe the statistics of kinks in the ground
state class $N$, we introduce the `wave function' $\Psi_N\{z\}$,
which equals the number of the minimal-energy kink configurations
in the lower half-plane for fixed positions of the $N$ kinks in
the uppermost chain represented by $\{z\} = (z_1,z_2,\ldots,z_N)$,
where $z_j = e^{2\pi i \frac{x_j}{L_x}}$ and $L_x$ is the chain
length. One can then add one more chain from above and obtain an
eigenvalue equation for the wave function
\begin{equation}
\lambda_N \Psi_N\{z\} = {\sum_{\{z'\}}}^\prime
\Psi_N\{z'\}, \label{eq:Psi}
\end{equation}
where the $\sum^\prime$ denotes the summation that
preserves the energetically favorable relative ordering of
kinks with the coordinates $\{z\}$ and $\{z'\}$ in two
neighboring chains. The solution of Eq.(\ref{eq:Psi}) is
the absolute value of the van der Monde determinant
\begin{equation}
\Psi_N(z_1,z_2,\ldots,z_N) \propto \prod_{i<j} |z_i -
z_j|, \label{eq:solution}
\end{equation}
It can also be written in the form of the Slater
determinant of $N$ plane waves $e^{i k_n x_j}$ ($n,j =
1,2,\ldots,N$) with the wave vectors $k_n$ taking values
in the Fermi sea $- k_F < k < + k_F$, where the Fermi wave
vector $k_F$ is related to the density of kinks $n =
\frac{N}{L_x}$ by $k_F = \pi n$. Thus,
Eq.(\ref{eq:solution}) is the ground-state wave function
of $N$ fermions in the chain with $L_x$ sites and the
kinks can be identified with the fermions appearing in the
transfer-matrix solution of the CFM \cite{PeschelEmery}.

\looseness = -1 The eigenvalue $\lambda_N$ in Eq.(\ref{eq:Psi}) is
related to the number of the ground states in this class, $W_N$,
by $W_N = \lambda_N^{L_y}$ ($L_y$ is the number of chains). In the
limit $L_x, L_y \rightarrow \infty$ and for a fixed kink density
$n$, $\lambda_N \sim e^{L_x S(n)} \frac{2\pi}{\sqrt{L_x \tan
\frac{\pi n}{2}}}$, where $S(n)$ is the ground-state entropy per
site
\begin{equation}
S(n) = \frac{1}{\pi} \int_0^{\pi n} d\phi \ln \left(2 \cos
\frac{\phi}{2} \right)\label{eq:entropy}.
\end{equation}
For $n = \frac{1}{3}$, at which $S(n)$ has its maximum,
Eq.(\ref{eq:entropy}) gives the value $\sim 0.323 k_B$
cited above.

\looseness = -1 The joint distribution function $P_N(\{z\})$ of
$N$ kinks in a chain is obtained by summing over all
minimal-energy configurations of kinks in the chains {\em both
below and above} this chain, which gives
\begin{equation}
P_N\{z\} = \Psi_N^2\{z\} \propto \prod_{i<j} |z_i -
z_j|^2. \label{eq:PN}
\end{equation}
The spin correlation function along chains $\langle
\sigma_x \sigma_0 \rangle$ = $(-)^x \langle (-)^{K(x)}
\rangle$, where $K(x)$ is the number of the kinks in the
interval $[0,x]$. Using Eq.(\ref{eq:PN}), the spin
correlator can be written in the form of the Toeplitz
determinant: $\langle \sigma_{x} \sigma_{0} \rangle =
(-)^{x} \det f_{nm}$, where $f_{nm} = \delta_{nm} -
\frac{2}{L_x} \frac{\sin \frac{\pi (n-m) x}{L_x}}{\sin
\frac{\pi (n-m) }{L_x}}$. For $1 \ll x \ll L_x$, the
determinant $\propto \frac{1}{\sqrt{x}} \cos k_F x$, so
that for $k_F = \frac{\pi}{3}$ we recover
Eq.(\ref{eq:correlator}). In general, the ground state
class with $N$ kinks/chain has an algebraic SDW order with
the wave vector $q = \pi(1 - n)$, which we interpret as
the $2k_F$-instability of the kink Fermi sea. For
Ising-type models with the discrete $Z_2$ symmetry such
SDW states are very unusual.

\looseness = -1 The number of ground states $W_N$ has a
sharp peak at $N_\ast = L_x /3$: $W_N \propto \exp \left[-
\mbox{const} \frac{L_y}{L_x}\left(N - N_\ast\right)^2
\right]$. Though the number of classes significantly
contributing to the CFM partition function stays finite in
the thermodynamic limit, all of them are essentially
identical copies of the class with $n = 1/3$, since the
corresponding SDW vectors deviate from $q = 2 \pi /3$ by
an amount $ \propto 1 / L_x$. Furthermore, the deviation
of the total ground state entropy per spin $S =
\frac{1}{L} \ln \sum_N W_N $, where $L = L_x L_y$ is the
number of spins, from the entropy of a single class
Eq.(\ref{eq:entropy}) at $n = \frac{1}{3}$ is $O(1/L)$.
Thus, in the thermodynamic limit it suffices to consider
only one class with 1 kink per 3 sites.

\looseness = -1 \noindent{\em Quantum model:} The transverse field
$h$ in Eq.(\ref{eq:quantum}) flips spins, resulting in hopping of
kinks along the chains, as well as in creation/annihilation of
kink-antikink pairs on neighboring chain sites. For $h \ll J$ the
hopping plays the dominant role, as it mixes degenerate classical
ground states within each class, whereas the kink-antikink pairs
cost energy $\sim 4J$ and only appear virtually, renormalizing the
kink dispersion: $ E(k) \approx - 2h \cos k + \frac{h^2}{J} \left(
\sin^2 k + \frac{1}{6} \right)$. Here $k$ is the kink wave vector,
the first term is due to the hopping of kinks on neighboring
sites, and the second term contains contributions from the
dressing of kinks and the ground state by one virtual
kink-antikink pair. For $h,T \ll J$ the positions of the dressed
kinks satisfy the same restrictions as in the classical ground
states and $h$ is the only relevant energy scale. In other words,
kinks form a quantum hard sphere liquid. The restricted motion of
kinks makes the quantum system more rigid than the classical one
and gives rise to phonon-like excitations with velocity $\propto
h$.

\looseness = -1 In the CFM kinks only crystallize at $T =
0$. At any nonzero temperature the algebraic order is
destroyed by dislocations in the kink crystal with energy
$2J$ (see Fig.\ref{fig:order}b), which in the classical
model are unbound. The latter leads to the asymptotic
behavior of the spin correlation function  at $T \ll J$
\cite{Stephenson}, given by the right-hand side of
Eq.(\ref{eq:correlator}) multiplied by $e^{-r_{ij}/\xi}$,
where $\xi = e^{2 \beta J}$ is the average distance
between the dislocations. On the other hand, in the QFM
phonons result in two-dimensional Coulomb interactions
$U(r) \propto h \ln r$ between the dislocations separated
by a distance $r$, binding them into pairs at a finite
temperature $T_1 \propto h$, below which the spin
correlations decay algebraically \cite{Kosterlitz}. Upon
lowering temperature, the exponent $\eta$ (see
Eq.(\ref{eq:correlator})) decreases, as the kink crystal
becomes more rigid, which ultimately leads to the pinning
of the kink crystal by the lattice at some temperature
$T_2 < T_1$, below which the phonons become gapped and the
spin ordering becomes long-ranged
\cite{PokrovskyUmin,Jose,HalperinNelson}. A cartoon of the
quantum kink crystal with 1 kink per 3 sites is shown in
Fig.\ref{fig:order}c. To gain kinetic energy each kink is
delocalized over one bond (such bonds are shown as dimers)
and the bonds are arranged in a way that ensures the
energetically favorable ordering of kinks. This state
corresponds to the SDW state with the wave vector $q =
\frac{2\pi}{3}$ (see Fig.~\ref{fig:order}d), in which the
spins antiferromagnetically ordered along the $z$ axis
form a bipartite hexagonal lattice, while the spins
located in the centers of the hexagons are oriented along
the transverse field, since the fields from their
neighbors add to zero.

\looseness = -1 Thus for weak transverse fields $h \ll J$,
the QFM describes a kink crystal that melts like a crystal
of adsorption atoms on a substrate lattice, i.e. the
melting is preceded by a depinning transition and the
solid phase is separated from the liquid phase by a
`floating crystal phase' with algebraic crystal order
\cite{PokrovskyUmin,Jose,HalperinNelson}. In the context
of the QFM this phase diagram was recently suggested by
Moessner et al.
\cite{MoessnerSondhiChandra,MoessnerSondhi}, who studied
the critical behavior of the QFM in the vicinity the
quantum critical point, i.e. in the opposite limit of
strong transverse fields $h \sim J$. Two-dimensional
crystals are also known to melt via a single first-order
transition due to proliferation of the boundaries between
degenerate crystal phases
\cite{HalperinNelson,Chui,Weber}. We compare these two
scenarios to our numerical results.

\begin{figure}[tbp]
\centering
\includegraphics[width=6cm]{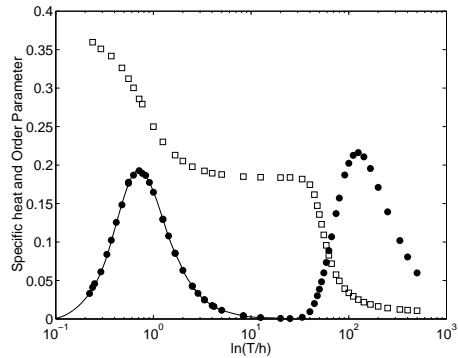}
\caption{{\protect\footnotesize The plot of specific heat
(circles) and the `order parameter' $m$ (squares) of the
QFM vs $h/T$, for $h \ll J$. The smooth line interpolates
the  part of $c(h/T)$ due to quantum superpositions of
classical ground states.}} \label{fig:small}
\end{figure}

\looseness = -1 \noindent{\em Numerical results:} We performed
Monte Carlo simulations of the QFM using the continuous time
algorithm \cite{Prokof'ev}, calculating the temperature dependence
of the specific heat $c$ and the susceptibilities $\chi_q =
\langle M_{q}^2 \rangle / L$, where $M_q = \sum_j \sigma^z_j \cos
q x_j$ and $q$ is a multiple of $\frac{2 \pi}{L_x}$. A large value
of $\chi_q$ is a signature of the SDW state with the wave vector
$q$. We also calculate the finite system `order parameter' $ m =
\sqrt{\langle M_{2\pi/3}^2 \rangle} / L$.

\looseness = -1 The behavior of the model is most
spectacular for weak transverse fields, $h \ll J = 1$,
when one can see a clear difference between the classical
and quantum regimes. In Fig.\ref{fig:small} we plot the
specific heat and `order parameter' as a function of the
ratio $T/h$, varying over four decades and covering both
the classical region $h \ll T,J$ and the quantum region
$h,T \ll J$. The `classical points' are calculated at $h =
0.01$, while the `quantum data' is a collection for many
$(h,T)$ points, which fall on smooth curves when plotted
versus $T/h$, showing that in the quantum regime $h$ is
indeed the only relevant energy scale. The specific heat
has two maxima: one at $T \sim J$, which also exists in
the classical model, and another at $T \sim h$, due to
quantum superpositions of classical ground states. To show
that the huge degeneracy of the classical model is lifted
by a transverse field, we calculate the entropy release
related to the low-$T$ maximum, $\Delta S =
\int_0^{T_\ast}\!\!\frac{dT}{T}c$, where $h \ll T_\ast \ll
J$. The numerical integration that uses the smooth-curve
fit of the MC data (see Fig.\ref{fig:small}) gives $\Delta
S = 0.32 k_B$, in perfect agreement with the zero
temperature entropy of the classical model. The `classical
maximum' corresponds to the disappearance of defects in
the kink crystal with the energy $\sim J$, which {\em in
finite systems} induces an algebraic spin ordering with
$\eta = \frac{1}{2}$, just as in the CFM at $T = 0$. The
latter is clear from the temperature dependence of the
`order parameter', which grows fast at $T \sim J$ and
stays constant at $h \ll T \ll J$. At $T \sim h$, the
`order parameter' grows again, reflecting the increasing
stiffness of the kink crystal due to the quantum motion of
kinks, which results in a decrease of the exponent $\eta$
and, perhaps, in the appearance of the long-range order.
The smooth temperature dependence of $c$ and $m$ is
evidence against a first-order transition and is
compatible with the two phase-transitions scenario, since
both transitions are expected to be of the
Kosterlitz-Thouless type \cite{Jose}. Furthermore, in a
finite system the transition at the upper critical
temperature $T_1 \sim h \ll J$, describing the dislocation
binding does not occur, as all dislocations disappear at
much higher temperature $\sim J$. The lower critical
transition is also rather difficult to identify, since the
algebraic order with $\eta = 1/9$ at $T = T_2 + 0$
\cite{Jose} is practically indistinguishable from the
long-range order at $T < T_2$.

\begin{figure}[tbp]
\centering
\includegraphics[width=8cm]{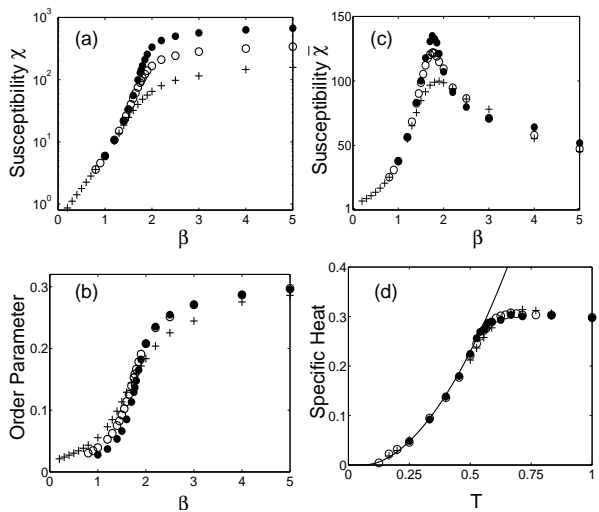}
\caption{{\protect\footnotesize The susceptibility $\chi$ (a), the
`order parameter' $m$ (b), and the susceptibility ${\bar \chi}$
(c) versus inverse temperature $\beta$ for $h = J =1$, $L_x = 96$
and three different values of $L_y$: $20$ (pluses), $40$ (open
circles), and $80$ (filled circles). Plotted in panel (d) is the
corresponding temperature dependence of the specific heat. The
solid line is the specific heat of two-dimensional phonons with a
gap $\sim 0.5$.}} \label{fig:opar1}
\end{figure}

\looseness = -1 In Figs.~\ref{fig:opar1}(a) and (b) we
plot $\chi \equiv \chi_{2\pi/3}$ and $m$ vs temperature,
for the strong transverse field $h = J = 1$, $L_x = 96$,
and $L_y = 20,40$, and $80$. For $\beta < 1.5$, the
susceptibility/spin $\chi$ is independent of $L_y$,
corresponding to a disordered phase, while for $\beta >
2$, the `order parameter' $m$ shows little size-dependence
for large $L_y$, indicating a long-range ordering. On the
basis of our data it is difficult to conclude whether the
intermediate region $1.5 < \beta < 2$ is the `algebraic'
phase with temperature-dependent $\eta$ or it is a
vicinity of a single transition. Note, that the specific
heat has a rather sharp kink at $\beta \sim 1.8$, at which
its behavior changes from approximately
temperature-independent to $T^2$-dependence, corresponding
to the specific heat of phonons in the two-dimensional
kink crystal (see Fig.~\ref{fig:opar1}(d)). Also the
susceptibility ${\bar \chi} = \sum_{q \neq 2\pi/3}
\chi_q$, describing SDWs with subdominant harmonics has a
peak at $\beta = 1.75$ (see Fig.~\ref{fig:opar1}(c)),
suggesting a single transition, which may be attributed to
a sudden loss of rigidity of the kink crystal.

\looseness = -1 \noindent{\em Topological spin glass?:} We
found that for $\beta J > 2$ it is effectively impossible
to bring a large system into thermal equilibrium, as it
`freezes' in one of the SDW states with $q \sim 2 \pi /
3$. This `glassy' behavior is related to the existence of
different ground state classes, which in the quantum case
transform into an array of energy valleys separated by
barriers. At low $T$ the barriers become impenetrable, as
the tunneling between neighboring classes requires
creating/annihilating a kink-antikink pair in all chains.
Does such an energy landscape lead to a spin-glass
behavior in the absence of quenched disorder and should
the `kink' in $c(T)$ at $T \sim 0.6$ (see
Fig.~\ref{fig:opar1}c) be interpreted as a spin-glass
transition? We believe that the answer is negative,
because (as in the CFM) the number of important energy
minima in the thermodynamics limit stays finite, all of
them describing essentially the same state. In our low-$T$
simulations we select the state with $q = 2 \pi /3$ by an
appropriate initial spin configuration.

\looseness = -1 In conclusion, we explained unusual
properties of the classical and quantum frustrated Ising
models by mapping them on a system of kinks, which behave
like hard spheres. We showed that the quantum spin
dynamics makes the kink crystal more rigid, resulting in a
long range ordering at low temperatures. We studied
numerically the temperature dependence of specific heat
and spin susceptibility in weak and strong transverse
fields.

This work is supported by the MSC$^{\mbox{\em plus}}$
program and the NSF grant DMR 0071767. We thank R.
Moessner and A. Tsvelik for useful discussions.

\end{document}